\documentstyle[prc,aps,psfig]{revtex}
\newcommand{\be}{\begin{eqnarray}}
\newcommand{\ee}{\end{eqnarray}}

\newcommand{\bi}{\begin{itemize}}
\newcommand{\ei}{\end{itemize}}
\begin{document}
\twocolumn[\hsize\textwidth\columnwidth\hsize
           \csname @twocolumnfalse\endcsname
\title{Quantum response of finite Fermi systems and the relation of Lyapunov exponent to transport coefficients}
\author{Klaus Morawetz}
\address{Fachbereich Physik, University Rostock, D-18055 Rostock,
Germany}
\maketitle
\begin{abstract}
Within the frame of kinetic theory a response function is derived for finite Fermi systems which includes dissipation in relaxation time approximation and a contribution from additional chaotic processes characterized by the largest Lyapunov exponent. A generalized local density approximation is presented including the effect of many particle relaxation and the additional chaotic scattering. For small Lyapunov exponents relative to the product of wave vector and Fermi velocity in the system, the largest Lyapunov exponent modifies the response in the same way as the relaxation time. Therefore the transport coefficients can be connected with the largest positive Lyapunov exponent in the same way as known from the transport theory in relaxation time approximation.
\end{abstract}
\vskip2pc]

The problem of irreversibility is one of the still open questions. Two approaches basically can be distinct. One approach considers the many particle theory as a suitable starting point to understand the increase of entropy as a result of many random collisions leading to irreversible kinetic equations like the Boltzmann equation. The other approach considers the theory of deterministic chaos with the characteristic measure of Lyapunov exponent to understand the occurrence of irreversibility. While the many particle approach can be easily extend to quantum systems the quantum chaos is still a matter of debate about the correct term. 

If both approaches describe some facet of irreversibility, what we will anticipate in the following, it should be possible to give relations between them. While the characteristic measure of many body effects is the relaxation time and the transport coefficients, the relevant measure for chaotic systems is the Lyapunov exponent as a measure of phase space spreading of trajectories. Considerable efforts have been made to connect the transport coefficients with the Lyapunov exponent 
\cite{evans,dorfman,dorfman1,C95}. In \cite{evans,C95} the fact, that the spreading of a small phase space volume is given by the sum of Lyapunov exponents $\delta V(t)=\delta V(0) \exp{(\sum \lambda_i)t}$, is used to give a relation between Lyapunov exponents and viscosity. This was possible to show with the help of the contact to a heat bath in the equation of motion ensuring constant internal energy. In \cite{dorfman,dorfman1} the relation between transport coefficients and Lyapunov exponents was presented in terms of Helfand's moments. The interlink was possible to establish by reinterpretation of the Helfand's moments as stochastic quantities such that the mean variance of the time derivatives represents just the transport coefficients. In \cite{DB97} the authors derive a density expansion of largest Lyapunov exponent for hard sphere gases from a generalized Lorentz-Boltzmann equation. This demonstrated the intimate relation between transport coefficients and dynamical quantities like the Lyapunov exponent.

Here we like to show that there exists a very simple connection between the concept of Lyapunov exponent and the dissipation leading to irreversibility for interacting Fermi systems. It will be shown that if the largest positive Lyapunov exponent is smaller than the product of Fermi velocity times wavelength in a Fermi system, the Lyapunov exponent appears in the same way as the relaxation time of the system. Therefore all expressions known from kinetic theory, expressing the transport coefficients in terms of the relaxation time, can be considered as an expression in terms of the Lyapunov exponent.

The concept of response of an interacting many body system starts conveniently
from the distribution function
      $f({\bf p,r},t)$ satisfying the appropriate kinetic equation, which by linearization yields the response to an external disturbance. First we discuss the quasiclassical response and generalize later to quantum response. The starting quasiclassical kinetic equation reads
\be
&&{\partial_t f}({\bf p},{\bf r},t)+\frac{{\bf p}}{m}\partial_{\bf r}f({\bf p},{\bf r},t)\nonumber\\
&&-\partial_{\bf r}(V_{\rm ind}({\bf r},t)+V_{\rm ext}({\bf r},t)) \partial_{\bf p}f({\bf p},{\bf r},t)
={f_0({\bf p},{\bf r})-f({\bf p},{\bf r},t) \over \tau}\nonumber\\
&&
                    \label{klassVlasov}
\ee
with the self-consistent mean-field potential given as a convolution between the two-particle interaction $V_0$ and the density $V_{\rm ind}=\int d{\bar {\bf r}} V_0({\bf r},{\bar {\bf r}}) n({\bar {\bf r}},t)$, the external disturbance $V_{\rm ext}$ and a typical relaxation time
$\tau$. The relaxation time approximation serves here as the simplest form of collision integral to describe dissipative processes by internal collisions of the particles. This leads to a natural chaotization and ergodicity of the system. 

Besides this chaotization by mutual collisions we want to discuss in the following how additional chaotic processes, e.g. caused by boundary conditions, surfaces etc., are influencing the response of the system to external perturbation $V_{\rm ext}$.

When the equation (\ref{klassVlasov}) is linearized with respect to the external perturbation, the selfconsistent potential $V_{\rm ind}$ gives a linear density contribution $\delta n$ via $\delta V_{\rm ind}=V_0\delta n$. Defining the total polarization function as the connection between induced density variation and external perturbation
\be
\delta n({\bf x},\omega)=\int d{\bf x}'\,\Pi({\bf x,x'},\omega)\;\delta V_{\rm ext}({\bf x'},\omega),
\label{5}
\ee
one finds the relation between the polarization function including the effect of the selfconsistent potential, $\Pi$, and the polarization without selfconsistent potential, $\Pi_{\tau}$, as
\be
\label{l}
&&\Pi({\bf x,x'})=\Pi_{\tau}({\bf x,x'})+\int \!\!\! d {\bar {\bf x}}d {\bar {\bar {\bf x}}} \,\Pi_{\tau}({\bf x,{\bar x}}) V_0({\bf {\bar x},{\bar {\bar x}}}) \Pi({\bf {\bar {\bar x}},x'}).\nonumber\\
&&
\ee
In other words it is sufficient to concentrate on the response function $\Pi_{\tau}$ to an external potential without selfconsistent potential $V_{\rm ind}$. The selfconsistent response $\Pi$ is then given by the solution of the integral equation (\ref{l}). The following derivation of $\Pi_{\tau}$ is adapted from \cite{kirschnitz}.

Introducing the Lagrange picture by following the trajectory ${\bf x}(t), {\bf p}(t)$ of 
a particle
\be
{d \over dt} {\bf x}&=&{{\bf p}\over m}\nonumber\\
{d \over dt} {\bf p}&=&-\partial_{\bf x} V_{\rm ext}
\label{Hamilton}
\ee
we linearize the kinetic equation
equation (\ref{klassVlasov}) according to $f({\bf x,p},t)=f_0({\bf x,p})+\delta f({\bf x,p},t) {\rm e}^{-t/\tau}$ and obtain
\be
\frac{d}{dt}\delta f({\bf x}(t),{\bf p}(t),t)= \partial_{\bf p} f_0\,\partial_{{\bf x}(t)}\,V_{\rm ext}.
\ee
This can be integrated to yield
\be
&&\delta f({\bf x,p},t) = \nonumber\\
&&-2m \!\! \int\limits_{-\infty}^0 \!\!\!\! dt'\!\!\int\limits_{-\infty}^{\infty} \!\!\! d{\bf x}'\frac{d}{dt'}\delta({\bf x'}-{\bf x}(t')) \; 
                {\partial f_0(p^2,{\bf x'}) \over \partial {p^2}}\,V_{\rm ext}({\bf x}',t+t').  
\nonumber \\
\ee
Integrating over ${\bf p}$, the density variation caused by varying the external potential is obtained as
\be
&&\delta n({\bf x},\omega)=
-2mg\int d{\bf x}'\int  \frac{d{\bf p}^3}{(2\pi\hbar)^3}\partial_{p^2}
                                         f_0(p^2,{\bf x'})\nonumber\\
&&\times\int\limits_{-\infty}^0 dt'e^{-it'
      (\omega +{i\over \tau})}V_{\rm ext}({\bf x'},\omega)\frac{d}{dt'}\delta({\bf x'-x}(t')) 
\label{4}
\ee
where $g$ denotes the spin-isospin degeneracy. Comparing the expression (\ref{4}) with 
the definition of the polarization function 
$\Pi_{\tau}$ in (\ref{5}) and (\ref{l}),
we are able to identify the polarization of finite systems including the relaxation time as
\be
\Pi_{\tau}({\bf x,x'},\omega)=\Pi_0({\bf x,x'},\omega+{i\over \tau})
\label{8}
\ee 
with
\be
\Pi_0({\bf x,x'},\omega)&=&- 2mg\int \frac{d{\bf p}^3}{(2\,\pi\,\hbar)^3}\partial_{p^2}\,
f_0(p^2,{\bf x'})\nonumber\\
&\times&\int\limits_{-\infty}^0 dt'\, e^{-i\,t'\,
\omega}\frac{d}{dt'}\delta({\bf x'-x}(t')). 
\label{9}
\ee

Further simplifications are possible if we focus on the ground state
$f_0(p^2)=\Theta(p_f^2-p^2)$ of the Fermi system. The modulus integration of momentum can 
be carried out and the Kirschnitz-formula 
\cite{kirschnitz,DM95} appears
\be
&&\Pi_0({\bf x,x'},\omega)= -\frac{m g p_f({\bf x})}{4 \pi^2 \hbar^3}\Bigg[ 
\delta({\bf x'}-{\bf x}(0))\nonumber\\
&&+\,i  \omega\int\limits_{-\infty}^0 dt'  e^{-i t' 
\omega}\int\limits_{}^{} 
\frac{d\Omega_{\bf p}}{4 \pi}\delta({\bf x'}-{\bf x}(t'))\Bigg]. 
\label{prewigner}
\ee
This formula represents the ideal free part and a contribution which arises 
by the trajectories ${\bf x}(t)$ averaged over the 
direction at the present time ${\bf n}_p p_f=m {\dot {\bf x}}(0)$. In principle, the knowledge of the evolution of all 
trajectories is necessary to 
evaluate this formula. Molecular dynamical simulations can perform this task but it requires an astronomical amount of memory to store all trajectories. Rather, we discuss two approximations which will give us more insight into the physical processes behind. 
First the most radical one shows how the local 
density approximation emerges. In the next one we consider the influence of 
chaotic scattering.

The local density approximation appears from (\ref{prewigner}) when  we perform two simplifications. Introducing Wigner 
coordinates ${\bf R}=({\bf x}+{\bf x'})/2$, ${\bf r}={\bf x}-{\bf x'}$ we have to assume
\begin{enumerate}
\item
gradient expansion
\be
p_f({\bf R}+\frac{{\bf r}}{2})\approx p_f({\bf R}) +{\cal O}(\partial_{\bf R})
\ee
\item
expansion of the trajectories to first order history 
\be
{\bf x}'-{\bf x}(t')\approx -{\bf r}-t' {\dot {\bf x}}+ {\cal O}(t'^2) =
          -{\bf r}-t' \frac{p_f}{m}  {\bf n}_p .\nonumber\\
&&
\label{rule}
\ee
\end{enumerate}
With these two assumptions we obtain from (\ref{prewigner}) after trivial 
integrations
\be
&&\Pi_0^{\rm LDA}({\bf q},{\bf R},\omega)=\int \!\! d{\bf r} \,{\rm e}^{-i {\bf q r}} \,\Pi_0^{\rm LDA}({\bf r},{\bf R},\omega)
\nonumber\\&&
=
-\frac{m s p_f({\bf R})}{4 \pi^2 \hbar^3} 
\Bigg\{
1+i k \int\limits_{0}^{\infty} dy  
{\rm e}^{ik y}\frac{\sin \displaystyle{y}}{\displaystyle{y}}\Bigg\}
                        \label{pre1}
\ee
where $\it{k}=m\omega/(q p_f({\bf R}))$. This can be further integrated with the help of
\be
&&\int\limits_0^{\infty} dy {\rm e}^{i k y} {\sin y\over y}= {\rm arctan} ({\rm Im} \, k-i {\rm Re} \, k)^{-1}\nonumber\\
&& =\left . 2 i\ln \left( 
\frac{1+k} {1-k} \right) + \pi \left[\mbox{sgn}\left(1+k
\right)+\mbox{sgn}\left(1-k \right) \right]\right |_{{\rm Im}\, k \to 0}\nonumber\\
&&
\ee
to yield the standard 
Lindhard result (\ref{po}) in the classical limit
\be
\Pi_0^{\rm inf}({\bf q},p_f,\omega)&=&-\frac{m g p_f}{4 \pi^2 \hbar^3} 
\Bigg\{
1-2 k\ln \left( 
\frac{1+k} {1-k} \right) \nonumber\\
&+& i k \pi \left[\mbox{sgn}\left(1+k
\right)+\mbox{sgn}\left(1-k \right) \right]\Bigg\}
\ee
where $\it{k}=m\omega/(q p_f)$.
We recognize the ground state result for infinite 
matter except that the Fermi 
momentum $p_f({\bf R})$ has to be understood as a local 
quantity corresponding to local densities so that we get with (\ref{8})
\be
\Pi_{\tau}^{\rm LDA}({\bf q},{\bf R},\omega)=\Pi_{0}^{\rm inf}({\bf q},p_f({\bf R}),\omega+{i\over \tau})\label{lda}.
\ee
For extensions beyond the local density approximation see \cite{DM95,DM90}.

Now we focus on the influence of an additional chaotic scattering
which will be caused e.g. by a surface boundary. In order 
to investigate this effect we add to the regular motion (\ref{rule}) a small 
irregular part $\Delta {\bf x}$ 
\be
{\bf x}'-{\bf x}(t')\approx -{\bf r}-t' \frac{p_f}{m}  {\bf n}_p +\Delta {\bf x}.
\label{rule2}
\ee
The irregular part of the motion we specify in the direction of the current
movement lasting a time $\Delta_t$ and given by
an exponential increase in phase-space controlled by the largest Lyapunov exponent 
$\lambda$. Therefore we can assume [$t'<0$]
\be
\Delta {\bf x}\approx {p_f {\bf n_p} \over m} \Delta_t \exp[-\lambda 
(t'-\Delta_t)]+\mbox{const.} \label{deltax}
\ee
Since we are looking for the largest Lyapunov exponent 
we can take (\ref{deltax}) 
at the maximum $\Delta_t=-1/\lambda$. 
Further, we require, that in the case of vanishing Lyapunov exponent we 
should regain the regular motion (\ref{rule}). We have for (\ref{rule2}) therefore
\be
{\bf x}'-{\bf x}(t')\approx -{\bf r}-\frac{p_f}{m}  
{\bf  n}_p \left [{1-\exp(-\lambda t') \over \lambda} \right ].
\label{rule3}
\ee
With this ansatz one derives from (\ref{prewigner}) instead of (\ref{pre1}) 
the result
\be
&&\Pi_{\lambda}({\bf q},{\bf R},\omega)=-\frac{m g p_f({\bf R})}{4 \pi^2 \hbar^3}\nonumber\\
&&\times \Bigg[1+i k\int\limits_{0}^{\infty} d y \frac{\sin 
\displaystyle{y}}{\displaystyle{y}} \left (1+ {k y \over \omega} \lambda\right )^{i \omega/\lambda-1}\Bigg],                                                       \label{pre2}
\ee
which for $\lambda\rightarrow 0$ resembles exactly (\ref{pre1}). The further 
integration could be given in terms of 
hypergeometric functions but this is omitted here.

With this formula together with (\ref{8}) and (\ref{l}) we have derived the main result of a polarization function including the influence of many 
particle effects and additional chaotic process 
characterized by the Lyapunov exponent $\lambda$.

For the condition 
\be
\lambda<< q v_F 
\label{cond}
\ee
with $v_f=p_f/m$ the Fermi velocity and $q$ the wave length we can 
use $\lim\limits_{x \to \infty}(1+a/x)^x=\exp(a)$ under the integral of (\ref{pre2}) and the final integration is 
performed with the result of (\ref{lda}) but a complex shift
\be
\Pi_{\lambda}^{\rm LDA}({\bf q,R},\omega)=\Pi_0^{\rm inf}({\bf q},p_f({\bf R}),\omega+i (\lambda+{1 \over \tau}))
\label{ldac}.
\ee
We obtain by this way just the known Matthiessen rule which states that the 
damping mechanisms are additive in the 
damping $\Gamma={1\over \tau}+\lambda$.

Next we discuss the quantum response function and we will see that all discussions outlined above can be straight forward applied to the quantum response function. Instead of the quasiclassical kinetic equation (\ref{klassVlasov}) we start now from the quantum kinetic equation \cite{KBA62}
\be
&&\partial_t f({\bf p,r},t)+{{\bf p}\over m}\partial_{\bf r} f({\bf p,r},t)-\frac 1 i \int d{\bf s} {d {\bf p}'\over (2 \pi \hbar)^3} \left [U({\bf r}+{{\bf s}\over 2})\right .\nonumber\\
&&\left .-U({\bf r} -{{\bf s}\over 2}) \right ] {\rm e}^{{i \over \hbar} {\bf s}({\bf p'}-{\bf p})}f({\bf p',r},t)={f_0({\bf p,r})-f({\bf p,r},t) \over \tau}
\nonumber\\
&&
\ee
with $U=V_{\rm ind}+V_{\rm ext}$. The gradient expansion in $U$ leads to first order the quasiclassical expression (\ref{klassVlasov}).
We follow now exactly the same linearization as above and introduce the Langrange picture. The trajectories are now described instead of (\ref{Hamilton}) by the following set
\be
{d \over dt} {\bf x}&=&{{\bf p}\over m}\nonumber\\
{\bf s} {d \over dt} {\bf p}&=&U({\bf r}+{{\bf s}\over 2})-U({\bf r}-{{\bf s}\over 2})
\label{Hamiltonq}
\ee
where the arbitrary vector ${\bf s}$ shows the infinite possibilities of trajectories by quantum fluctuations. The resulting polarization function for a finite quantum system reads now instead of (\ref{8})
\be
&&\Pi_0({\bf x,x'},\omega)=\frac{g}{\pi^2\hbar^3}\int {d {\bf p}\over (2 \pi \hbar)^3} \int d {\bf s} {\sin ({1 \over \hbar}{\bf sp})\over s}\nonumber\\
&&\times\partial_s \left ({\sin ( {1\over  \hbar} s p_f)\over s}\right )
\int\limits_{-\infty}^0 dt'  e^{-i t' 
\omega} \delta ({\bf x'}-{\bf x}(t')-{{\bf s}\over 2})
\label{prewignerq}.
\ee
Compared with (\ref{9}) we see that due to quantum fluctuations an additional integration ${\bf s}$ appears. Eq. (\ref{prewignerq}) is the quantum generalization of the quasiclassical Kirschnitz formula (\ref{prewigner}) for the response function in finite systems.

Applying now the same gradient approximation (\ref{rule}) we derive from (\ref{prewignerq}) with the help of 
\begin{eqnarray}
&&i \int d{\bf s}  {\sin({1\over\hbar}{\bf  sp})\over s}\partial_s \left ({\sin ({1\over\hbar} s p_f)\over s}\right ){\rm e}^{i\frac 1 2 {\bf q s}}=\nonumber\\
&&\pi^2\hbar^3 \left [\Theta(p_f^2-({\bf p}-{{\bf q}\over 2})^2)-\Theta(p_f^2-({\bf p}+{{\bf q}\over 2})^2)\right ]
\end{eqnarray}
the quantum Lindhard result
\be
&&\Pi_0({\bf q,R},\omega)=\nonumber\\
&&g \int\!\!\!\! {d {\bf p}\over (2 \pi \hbar)^3} {\Theta(p_f^2({\bf R})\!-\!({\bf p}\!-\!{\hbar {\bf q}\over 2})^2)\!-\!\Theta(p_f^2({\bf R})\!-\!({\bf p}+{\hbar {\bf q}\over 2})^2)\over \hbar \omega-{\hbar {\bf p q}\over m}+i \epsilon}\nonumber\\
&&
\label{po}
\ee
in local density approximation.

The ansatz about additional chaotic processes (\ref{rule3}) leads then to exactly the same expression (\ref{ldac}) under the condition (\ref{cond}) but with the quantum response (\ref{po}) instead of $\Pi_0^{\rm inf}$.

 We like to point out that this result has far reaching consequences. With the assumption (\ref{cond}) we have shown by this way that the linear response behavior is the same if dissipation comes from the relaxation time via collision processes in many - particle theories or from the concept of chaotic processes characterized by the Lyapunov exponent. We can therefore state that for small Lyapunov exponent compared to the product of wave length and Fermi velocity in a many particle system, the largest Lyapunov exponent behaves like the relaxation time in the response function.

Since the transport theory is well worked out to calculate the transport coefficients in relaxation time approximation we can express by this way the transport coefficients in terms of the Lyapunov exponent alternatively. This illustrates the mutual equivalence of the concept of Lyapunov exponent and dissipative processes in many-particle theories. 

Pavel Lipavsk\'y and V\'aclav \v Spi\v cka are thanked for many enlightening 
discussions and A. Dellafiore for bringing the Kirshnitz formula to my attention.


\begin{thebibliography}{1}

\bibitem{evans}
D.~J. Evans, E. Cohen, and G.~P. Morriss, Phys.Rev.A {\bf 42},  5990  (1990).

\bibitem{dorfman}
J. Dorfman and P. Gaspard, Phys.Rev.E {\bf 51},  28  (1995).

\bibitem{dorfman1}
P. Gaspard and J. Dorfman, Phys.Rev.E {\bf 52},  3525  (1995).

\bibitem{C95}
E. Cohen, Physica A {\bf 213},  293  (1995).

\bibitem{DB97}
J. Dorfman and P. Gaspard,  (North-Holland, Amsterdam, 1997), Vol.~1/2, p.\ 12,
  proceedings of the Euroconference on The Microscopic Approach to Complexity
  in Non-Equilibrium Molecular Simulations CECAM, 15-19 July 1996.

\bibitem{kirschnitz}
D. Kirzhnitz, Y. Lozovik, and G. Shpatakovskaya, Usp. Fiz. Nauk {\bf 117},  3
  (1975).

\bibitem{DM95}
A. Dellafiore, F. Matera, and D.~M. Brink, Phys. Rev. A {\bf 51},  914  (1995).

\bibitem{DM90}
A. Dellafiore and F. Matera, Phys. Rev. A {\bf 41},  4958  (1990).

\bibitem{KBA62}
L. Kadanoff and G. Baym, {\em Quantum Statistical Mechanics} (Addison-Wesley,
  New York, 1962).

\end{thebibliography}

\end{document}